# Depth-resolved profile of the interfacial ferromagnetism in $CaMnO_3/CaRuO_3$ superlattices

J. R. Paudel[1], A. Mansouri Tehrani[2], M. Terilli[3], M. Kareev[3], J. Grassi[1], R. K. Sah[1], L. Wu[3], V. N. Strocov[4], C. Klewe[5], P. Shafer[5], J. Chakhalian[3], N. A. Spaldin[2], and A. X. Gray[1]

[1] *Physics Department, Temple University, Philadelphia, Pennsylvania 19122, USA*

[2] *Materials Theory, ETH Zurich, Wolfgang-Pauli-Strasse 27, CH-8093 Zürich, Switzerland*

[3] *Department of Physics and Astronomy, Rutgers University, Piscataway, New Jersey 08854, USA*

[4] *Swiss Light Source, Paul Scherrer Institute, 5232 Villigen, Switzerland*

[5] *Advanced Light Source, Lawrence Berkeley National Laboratory, Berkeley, California 94720, USA*

**email: axgray@temple.edu*

## ABSTRACT

Emergent magnetic phenomena at interfaces represent a frontier in materials science, pivotal for advancing technologies in spintronics and magnetic storage. In this letter, we utilize a suite of advanced X-ray spectroscopic and scattering techniques to investigate emergent interfacial ferromagnetism in oxide superlattices comprised of antiferromagnetic $CaMnO_3$ and paramagnetic $CaRuO_3$. Our findings challenge prior theoretical models by demonstrating that the ferromagnetism extends beyond the interfacial layer into multiple unit cells of $CaMnO_3$ and exhibits an asymmetric profile. Complementary density functional calculations reveal that the interfacial ferromagnetism is driven by the double exchange mechanism, facilitated by charge transfer from Ru to Mn ions. Additionally, defect chemistry, particularly the presence of oxygen vacancies, likely plays a crucial role in modifying the magnetic moments at the interface, leading to the observed asymmetry between the top and bottom $CaMnO_3$ interfacial magnetic layers. Our findings underscore the potential of manipulating interfacial ferromagnetism through point defect engineering.

## INTRODUCTION

The control of magnetic properties in oxide superlattices has attracted significant research interest due to their potential applications in spintronics [1-5]. Specifically, the stabilization and control of interfacial ferromagnetic ground states in material systems composed of two nonferromagnetic materials hold significant importance from both a fundamental and technological perspective [6-9].

The earliest and perhaps the best-known examples of such a material systems are oxide superlattices comprised of antiferromagnetic $CaMnO_3$ and paramagnetic $CaRuO_3$ layers, which have been extensively studied for their ferromagnetic properties [9-16]. In a pioneering study, Takahashi *et al.* [9] demonstrated a ferromagnetic transition at approximately 95 K, localized near the interface region. The magnetization and magnetoconductance of the superlattice remained constant and independent of the varying thickness, indicating the crucial role of the interface in the observed ferromagnetic-like behavior.

A subsequent theoretical study [10] explained this experimental observation by finding an exponential leakage of metallic Ru $3d$ $e_g$ electrons across the interface into the insulating $CaMnO_3$. This charge transfer was shown to stabilize the ferromagnetic state at the interface through ferromagnetic Anderson-Hasegawa double exchange [17,18], which competed with the antiferromagnetic superexchange in bulk $CaMnO_3$ to form of a one-unit-cell-thick ferromagnetic interfacial $CaMnO_3$ layer. The calculations also indicated minimal electron penetration beyond the interfacial layer, explaining the bulk antiferromagnetism in the remaining $CaMnO_3$.

Subsequent experiments yielded conflicting results regarding the size of the ferromagnetic unit cell. One experimental investigation, using a combination of spectroscopic probes, demonstrated that the aforementioned ferromagnetic polarization extends 3-4 unit cells (u.c.) into $CaMnO_3$, surpassing the one-unit-cell limit and suggesting the presence of magnetic polarons at the interface [12]. However, another study, employing polarized neutron reflectivity, revealed that interfacial ferromagnetism is indeed confined to only one unit cell of $CaMnO_3$ at each interface [14]. In addition to this, it has been suggested that the magnitudes of the interfacial Mn magnetic moments could be modulated by changing the symmetry of oxygen octahedra connectivity at the boundary, thus proposing the tuning of interfacial symmetry as a new route to control emergent interfacial ferromagnetism. [16].

In this letter, we present an in-depth analysis of interfacial ferromagnetism in $CaMnO_3/CaRuO_3$ superlattices, leveraging advanced synchrotron-based resonant X-ray reflectivity (XRR) techniques and density functional calculations to explore the magnetic properties at the interface of the two materials. We demonstrate that the ferromagnetic signal extends significantly beyond a single unit cell into the $CaMnO_3$, challenging previous theoretical models. Density functional calculations confirm that interfacial ferromagnetism is driven by a double exchange mechanism, facilitated by charge transfer from Ru to Mn across the interface, and show that oxygen vacancies change the sizes of the Mn magnetic moments. Detailed fitting of the $q_z$-dependent X-ray magnetic circular dichroism (XMCD) asymmetry spectra reveals pronounced magnetic asymmetry between the top and bottom magnetic interfaces. Our findings suggest that the presence of point defects, particularly oxygen vacancies, significantly influences the magnitude of the magnetic moments, offering a potential method to manipulate interfacial ferromagnetism in oxide superlattices for advanced spintronic applications.

**RESULTS**

A high-quality epitaxial superlattice consisting nominally of [4 u.c. $CaMnO_3$ / 4 u.c. $CaRuO_3$]×10 was synthesized on a single-crystalline $LaAlO_3$ (001) substrate using pulsed laser interval deposition [19]. In-situ monitoring of layer-by-layer growth was conducted using reflection high-energy electron diffraction (RHEED). The coherent epitaxy, crystallinity, and layering of the superlattice were verified through ex-situ X-ray diffraction spectroscopy (XRD) and X-ray reflectivity (XRR). To confirm the correct elemental layering of the superlattice, standing-wave photoemission spectroscopy (SW-XPS) [20] measurements were carried out at the soft-X-ray ARPES endstation [21] of the high-resolution ADRESS beamline at the Swiss Light Source [22]. The correct chemical composition was confirmed using bulk-sensitive hard X-ray photoelectron spectroscopy (HAXPES) measurements [23] with a laboratory-based spectrometer. Furthermore, synchrotron-based soft X-ray resonant and non-resonant reflectivity measurements, described in detail later in this article (Figures 1 and 2), were used to determine the individual layer thicknesses and assess the interface quality. The characterization results of XRD, XRR, SW-XPS, and HAXPES are presented in Figures S1, S2, and S3 of the Supporting Information.

To derive the detailed X-ray optical depth profile as well as the element-specific (Mn) magneto-optical profile of the superlattice we utilized polarization-dependent soft X-ray resonant and non-resonant reflectivity at the high-resolution ($\Delta E \approx 100$ meV) Magnetic Spectroscopy beamline 4.0.2 at the Advanced Light Source [24]. All measurements were carried out in an applied in-plane magnetic field of 0.1 T and at the sample temperature of 20 K, which is well below the reported $T_c$ (~95 K) for this system [9].

Figure 1a shows circular polarization-dependent XRR energy scans across the Mn $L_3$ and $L_2$ absorption edge carried out at a constant value of momentum transfer $q_z$ in specular X-ray incidence geometry. The XMCD difference ($I_{LCP}$ - $I_{RCP}$) and percent magnetic asymmetry ($I_{LCP}$ - $I_{RCP}$ / $I_{LCP}$ + $I_{RCP}$) are shown in the bottom panels and indicate ferromagnetism on the Mn sites. These data, measured in specular reflectivity mode, can be compared to the standard X-ray absorption (XAS) and XMCD spectra recorded in the total electron yield (TEY) mode of acquisition on the same sample, as shown in Figure S4a of the Supporting Information. These spectra show excellent agreement with the prior XAS studies of the $CaMnO_3$/$CaRuO_3$ superlattices [12,16]. Furthermore, as in these prior studies, they reveal fine spectral features attributed to the $Mn^{3+}$ and $Mn^{4+}$ cations. This suggests a mixed Mn valence state in $CaMnO_3$, which is required for the $Mn^{3+}$-$Mn^{4+}$ ferromagnetic double exchange interaction [18].

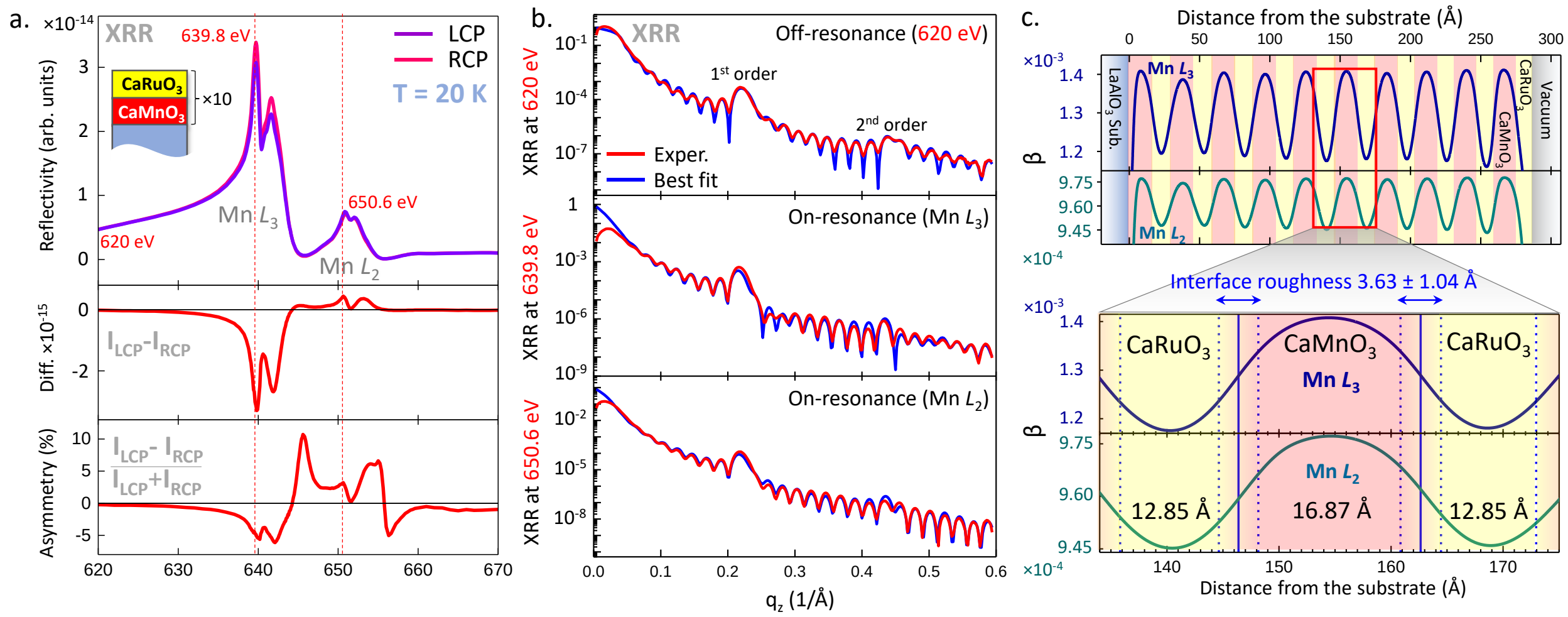

**Figure 1.** **a.** Upper panel: circular polarization-dependent XRR energy scans across the Mn $L_3$ and $L_2$ absorption thresholds. The measurements were carried out at a constant value of momentum transfer $q_z$ and at the temperature of 20 K. XRR XMCD difference ($I_{LCP}$-$I_{RCP}$) and magnetic asymmetry ($I_{LCP}$-$I_{RCP}$)/($I_{LCP}$+$I_{RCP}$) are shown in the lower panels. Three key photon energies corresponding to the non-resonant excitation (620 eV), the Mn $L_3$ peak (639.8 eV), and the Mn $L_2$ peak (650.6 eV) are marked with red dashed lines. **b.** Momentum-dependent XRR spectra and the best fits to the experimental data measured at the three above-mentioned photon energies. Self-consistent fitting of the data yields a detailed optical absorption coefficient β profile of the sample, shown in **c.**, with the extracted layer thicknesses of 12.85 Å ($CaRuO_3$) and 16.87 Å ($CaMnO_3$), as well as the average interface roughness (chemical interdiffusion) of 3.63 ± 1.04 Å.

To derive the detailed X-ray optical depth profile of the superlattice, we selected three photon energies corresponding to the off-resonant (620 eV) and resonant (Mn $L_3$ at 639.8 eV and $L_2$ at 650.6 eV) conditions and carried out $q_z$-dependent specular XRR scans that are shown in Figure 1b (red curves). The above-mentioned photon energies were selected by identifying the strongest peaks in the fixed-$q_z$ X-ray reflectivity and XMCD spectra (Fig. 1a). It is important to note that the $q_z$-dependent specular XRR spectra shown in Figure 1b span a wide range of $q_z$ (0 - 0.6 1/Å), encompassing both the first-order and second-order Bragg conditions (at ~0.22 and ~0.43 1/Å, respectively) and, therefore, contain detailed depth-resolved information on both the layering and the interfacial structure of the sample [25].

The $q_z$-dependent specular XRR spectra shown in Figure 1b (red curves) were fitted self-consistently with the XRR analysis program ReMagX [26], which uses an algorithm based on the Parratt formalism [27] and the Névot–Croce interdiffusion approximation [28]. For the off-resonant spectrum fitting, only the thicknesses of the $CaMnO_3$ and $CaRuO_3$ layers and the interdiffusion lengths between them were allowed to vary. The resonant X-ray optical constants required for the calculations were obtained by performing a Kramers-Kronig analysis of the XAS data and later optimized during the fitting of the XRR spectra. The blue-colored spectra shown in Figure 1b depict the best theoretical fits to the experimental data, demonstrating exceptional agreement in terms of the amplitudes of all features, as well as their relative phases and shapes.

A self-consistent X-ray optical profile of the superlattice resulting from the fitting of the three $q_z$-dependent specular XRR spectra is shown in Figure 1c. The profile is represented as the depth-dependent (x-axis) variation of the absorption coefficient β of the superlattice at the photon energies corresponding to the Mn $L_3$ (blue curve) and Mn $L_2$ (green curve) absorption edges. The maxima in such element-selective (Mn) absorption profiles correspond to the depth-resolved positions of the $CaMnO_3$ layers and the minima to the positions of the $CaRuO_3$ layers, where the Mn element is absent.

The lower part of Figure 1c shows a magnified view of the typical X-ray optical profile centered around a $CaMnO_3$ layer located approximately midway through the superlattice. The individual layer thicknesses obtained from the X-ray optical fitting are 12.85 Å for $CaRuO_3$ and 16.87 Å for $CaMnO_3$. These values correspond to approximately 3.5 and 4.5 primitive cubic unit cells of $CaRuO_3$ and $CaMnO_3$, respectively, using the lattice constants from prior studies [10,20,29]. However, we also find that the average interface roughness (interdiffusion) is 3.63 ± 1.04 Å, which corresponds to approximately one primitive cubic unit cell of a typical perovskite oxide. Therefore, although the average thicknesses of the individual $CaRuO_3$ and $CaMnO_3$ slabs in the superlattice may deviate slightly from the nominal 4 u.c./4 u.c. structure, these deviations are within the uncertainty introduced by the interfacial roughness. Note that the total extracted period of the superlattice is 29.72 Å, which corresponds exactly to the nominal value of 8 primitive cubic unit cells and is quantitatively consistent with both the lab-based XRR and synchrotron-based SW-XPS characterization, shown in the Supporting Figures S1 and S2, respectively. Another source of deviations in the calculated thicknesses of the individual layers can arise from minor inaccuracies in the resonant X-ray optical properties of the layers used during the fitting, because the X-ray optical constants vary drastically in the vicinity of the resonances (Mn $L_{2,3}$).

Thus, we have demonstrated that the use of polarization-averaged $q_z$-dependent specular XRR measurements in conjunction with X-ray optical modeling enables the determination of the X-ray optical profile of our $CaMnO_3/CaRuO_3$ superlattice, which also corresponds to the chemical/structural profile due to the use of element-specific (Mn) resonant photon energies. Building on this finding, next we utilized the extracted chemical/structural profile as an input in the model for fitting the $q_z$-dependent XMCD asymmetry ($I_{LCP}$ - $I_{RCP}$ / $I_{LCP}$ + $I_{RCP}$) spectra shown in Figure 2a. Since these magnetic asymmetry spectra are obtained via a simple mathematical operation from the same reflectivity data that was used for the chemical/structural analysis, this method self-consistently constrains the model, thus allowing for sensitive determination of the depth-resolved magneto-optical profile.

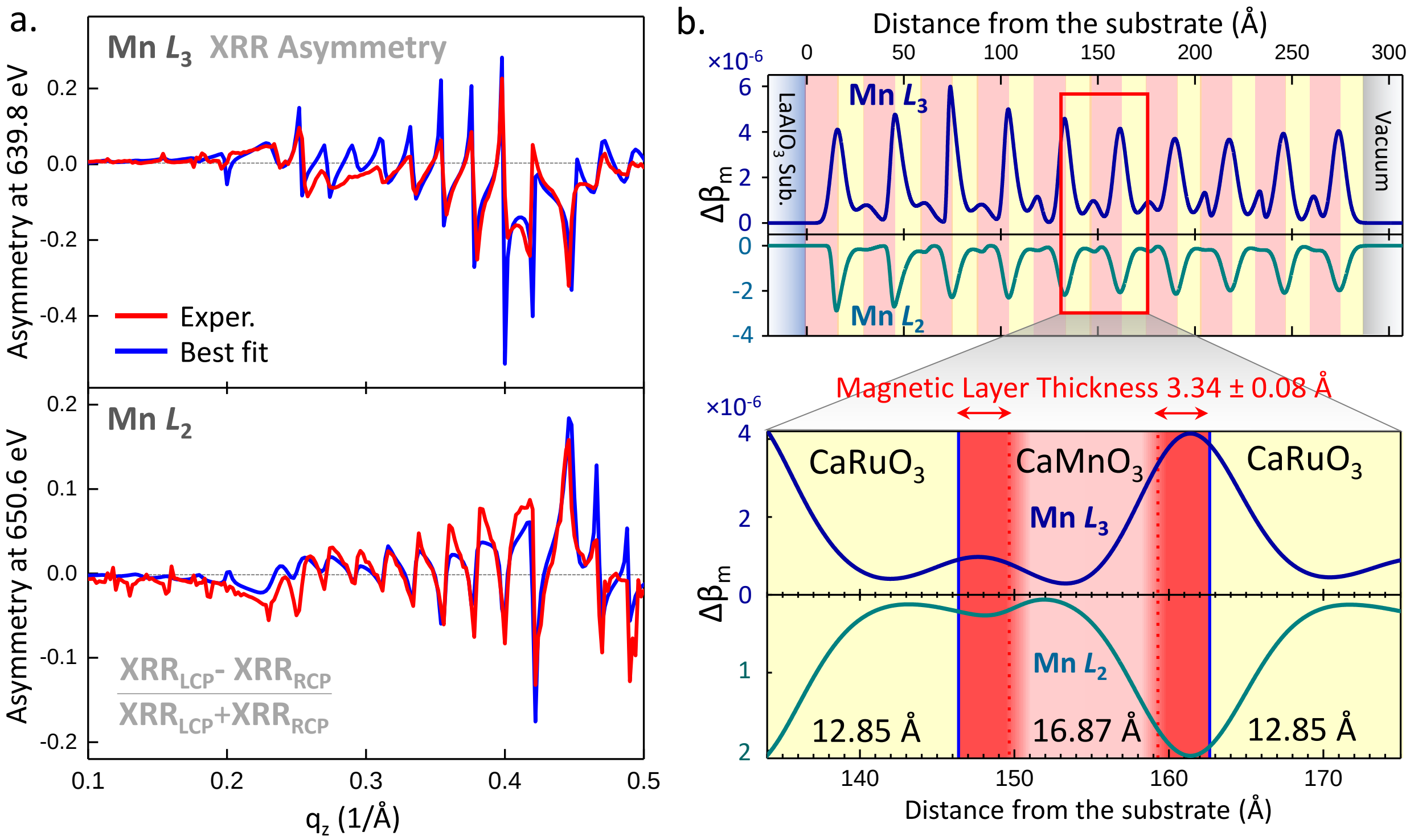

**Figure 2.** **a.** The $q_z$-dependent XMCD asymmetry spectra and the best fits to the experimental data measured at the resonant photon energies of the Mn $L_3$ (639.8 eV) and Mn $L_2$ (650.6 eV) XRR peaks. Self-consistent fitting of the data yields the detailed magneto-optical profile of the sample shown in b. **b.** Depth-resolved magneto-optical profile given by the modulation of the magnetic dichroism of the X-ray absorption coefficient $\Delta\beta_m$. The expanded region in the lower panel reveals an asymmetry in the magnetic moment at the top and bottom $CaMnO_3$ interfaces. The interfacial ferromagnetic $CaMnO_3$ layer thickness for both the top and bottom interfaces is estimated to be 3.34 ± 0.08 Å, with an additional magnetic interdiffusion length of 2.41 ± 1.14 Å into the bulk-like central region of the $CaMnO_3$ film and the adjacent $CaRuO_3$ layer.

We utilized the data collected at the photon energies of both Mn $L_3$ (top panel) and $L_2$ (bottom panel) edges, in order to further constrain the fitting. As previously, the experimental data are shown using red curves, with the optimal fits in blue. The only three parameters that were allowed to vary in the model were the thickness and roughness of the interfacial magnetic layer and the X-ray optical constant $\Delta\beta_m$, which quantifies the magnitude of the modulation of the magnetic dichroism of the X-ray absorption coefficient β. It is also important to note that the use of $q_z$-dependent XMCD asymmetry spectra significantly enhances the sensitivity of the fitting process due to the intricate spectral lineshapes, as depicted in Figure 2a, and the presence of numerous sharp modulations with varying amplitudes and shapes across the entire $q_z$ range. This improvement is in contrast to the traditional use of unnormalized $q_z$-dependent XMCD difference ($I_{LCP}$ - $I_{RCP}$) spectra, as commonly seen in similar studies.

The resultant magneto-optical profiles characterized by the thickness-dependent modulations of the values of $\Delta\beta_m$ at the resonant energies of the Mn $L_3$ (positive values, shown in blue) and Mn $L_2$ (negative values shown in green) are depicted in Figure 2b. The opposite signs are in agreement with the traditional convention for representing XMCD signals at the $L_3$ and $L_2$ edges. The difference in the amplitudes between the Mn $L_3$-derived and Mn $L_2$-derived profiles is also consistent with the expected differences in the XMCD signal at these two absorption edges (see Figure S4 in the Supporting Information).

The expanded region of Figure 2b shows the detailed magneto-optic profile of the $CaMnO_3$ layer, as well as the two adjacent $CaRuO_3$ layers, in the central region of the superlattice. The most striking feature of the profile is the several-fold (×5.5) asymmetry between signals at the bottom ($CaRuO_3$/$CaMnO_3$) and the top ($CaMnO_3$/$CaRuO_3$) interfaces, which will be discussed shortly. It is clear that the maxima of the magnetic signal are centered almost perfectly in the interfacial unit cells of $CaMnO_3$. The thickness of the magnetic layer is calculated to be 3.34±0.08 Å, which corresponds to approximately one primitive cubic unit cell of $CaMnO_3$. However, significant broadening of the magnetic signal, characterized by the Névot–Croce-type interdiffusion [28] with a characteristic width of 2.41±1.14 Å, is present on both sides of the magnetic layer. Therefore, although the ferromagnetism is clearly strongest in the interfacial unit cell of $CaMnO_3$, the total extent of the ferromagnetic signal in the sample is on the order of 2-3 cubic unit cells. This finding bridges the apparent discrepancies between studies that observe (or predict) interfacial ferromagnetism confined to a single interfacial unit cell of $CaMnO_3$ [10,14] and those showing that it extends several unit cells away from the interface [12].

In order to confirm the observed difference (asymmetry) between the magnitudes of the magnetic signal at the top and bottom $CaMnO_3$ interfaces, we first tried a different theoretical fitting, and then a different experimental technique. For the theoretical verification, we repeated the fitting of the $q_z$-dependent XMCD asymmetry spectra using the same model, but with an additional constraint of having the magnitudes of $\Delta\beta_m$ be the same for both interfaces. This modification resulted in a drastic deterioration in the quality of the fit (Figure S5 in the Supporting Information).

For the experimental verification, we used XAS/XMCD in the TEY detection mode to study a superlattice sample that was synthesized in the same batch as the first superlattice but terminated with reversed layers, so it terminated with the $CaMnO_3$ layer, instead of $CaRuO_3$. The TEY is a more surface-sensitive modality of XAS, with an average probing depth of 2-5 nm, decaying exponentially from the surface into the bulk [30,31]. Therefore, the XMCD measurement of the original $CaMnO_3$/$CaRuO_3$ sample (terminated with $CaRuO_3$) is most sensitive to the $CaMnO_3$/$CaRuO_3$ (or 'top' type) interface. Conversely, measurement of the $CaRuO_3$/$CaMnO_3$ sample (terminated with $CaMnO_3$) is most sensitive to the $CaRuO_3$/$CaMnO_3$ (or 'bottom' type) interface. The measurements reveal a significantly weaker (×2.7) magnetic signal for the $CaRuO_3$/$CaMnO_3$ ('bottom' type) interface than for the $CaMnO_3$/$CaRuO_3$ ('top' type), qualitatively consistent with our reflectivity measurements (Figure S4 of the Supporting Information). We speculate that, due to the exponential decay of the signal with depth, we still see some contribution to the total magnetic signal from the lower ($CaMnO_3$/$CaRuO_3$) interface, resulting in a weaker suppression of the depth-averaged magnetic signal (×2.7) as compared to that obtained using the depth-resolved XRR measurements (×5.5).

**DISCUSSION**

To shed light on the origin of the ferromagnetism at the interface, we performed DFT calculations of the structure and electronic properties of $CaMnO_3$/$CaRuO_3$ superlattices (see Supporting Information for details) [32,33]. In a first step, we calculated the energy for three magnetic configurations – the entire $CaMnO_3$ slab set to its bulk G-type antiferromagnetic (AFM) structure, the interfacial $CaMnO_3$ layers set to be ferromagnetic (FM) with the middle layers constrained to G-type AFM, and the entire $CaMnO_3$ slab set to be FM, re-relaxing the structure in each case. We found that the lowest energy arrangement has the G-type AFM ordering of the bulk in the central region of the $CaMnO_3$, with FM favored at the interface, consistent with our measurements; the relative energies are shown in Fig. 3b. Interestingly, interfacial FM is so strongly favored, that it is lower energy for the entire $CaMnO_3$ slab to adopt the FM configuration than for it all to have the AFM configuration of the bulk.

Having established that our calculations reproduce our measured interfacial magnetism, we next turn to understanding its origin. To this end, we calculated the layer resolved transition-metal Bader charges, magnetic moments, and densities of states; our results are shown as a function of layer number in Figures 3c, 3d and 3e respectively, with Ru values indicated in black and Mn values in blue. Our calculated Bader charges (Fig. 3c) indicate a charge transfer from the Ru layers to the Mn layers at the interface, consistent with the increased interfacial local Mn magnetic moment (Fig. 3d) and the metallic partial density of states (Fig. 3e). Therefore, our calculations point to a double exchange mechanism driven by interfacial metallicity as the origin of the ferromagnetism, as proposed in Refs. [8] and [10].

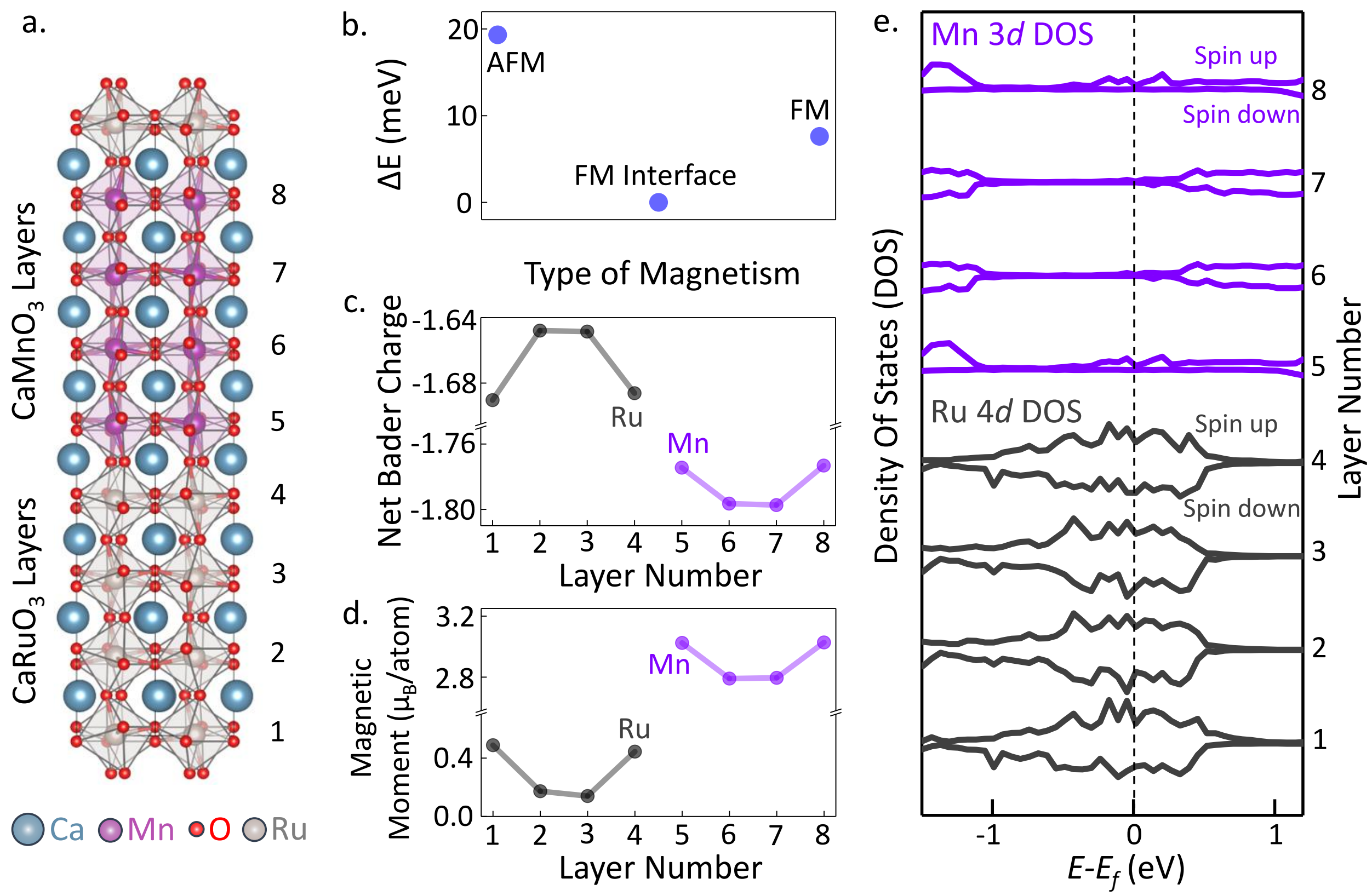

**Figure 3.** **a.** Crystal structure of the $CaRuO_3$/$CaMnO_3$ supercell (Ca in blue, Mn in magenta, O in red, Ru in grey) after ionic relaxation. **b.** Calculated energy differences between various magnetic states of the $CaMnO_3$ layers within the supercell: AFM denotes the entire 4-unit cell slab in an antiferromagnetic state; FM Interface indicates that only one unit cell at the interface exhibits ferromagnetism while the remaining bulk retains a bulk-like antiferromagnetic state; FM represents the entire $CaMnO_3$ slab in a ferromagnetic state. **c.** Net Bader charges for the individual layers of $CaRuO_3$ and $CaMnO_3$ in the supercell. **d.** Layer-resolved magnetic moments per atom for the Ru atoms in $CaRuO_3$ and the Mn atoms in $CaMnO_3$. **e.** Partial spin-projected densities of states for the Ru 4*d* states in the $CaRuO_3$ layers and for the Mn 3*d* states in the $CaMnO_3$ layers.

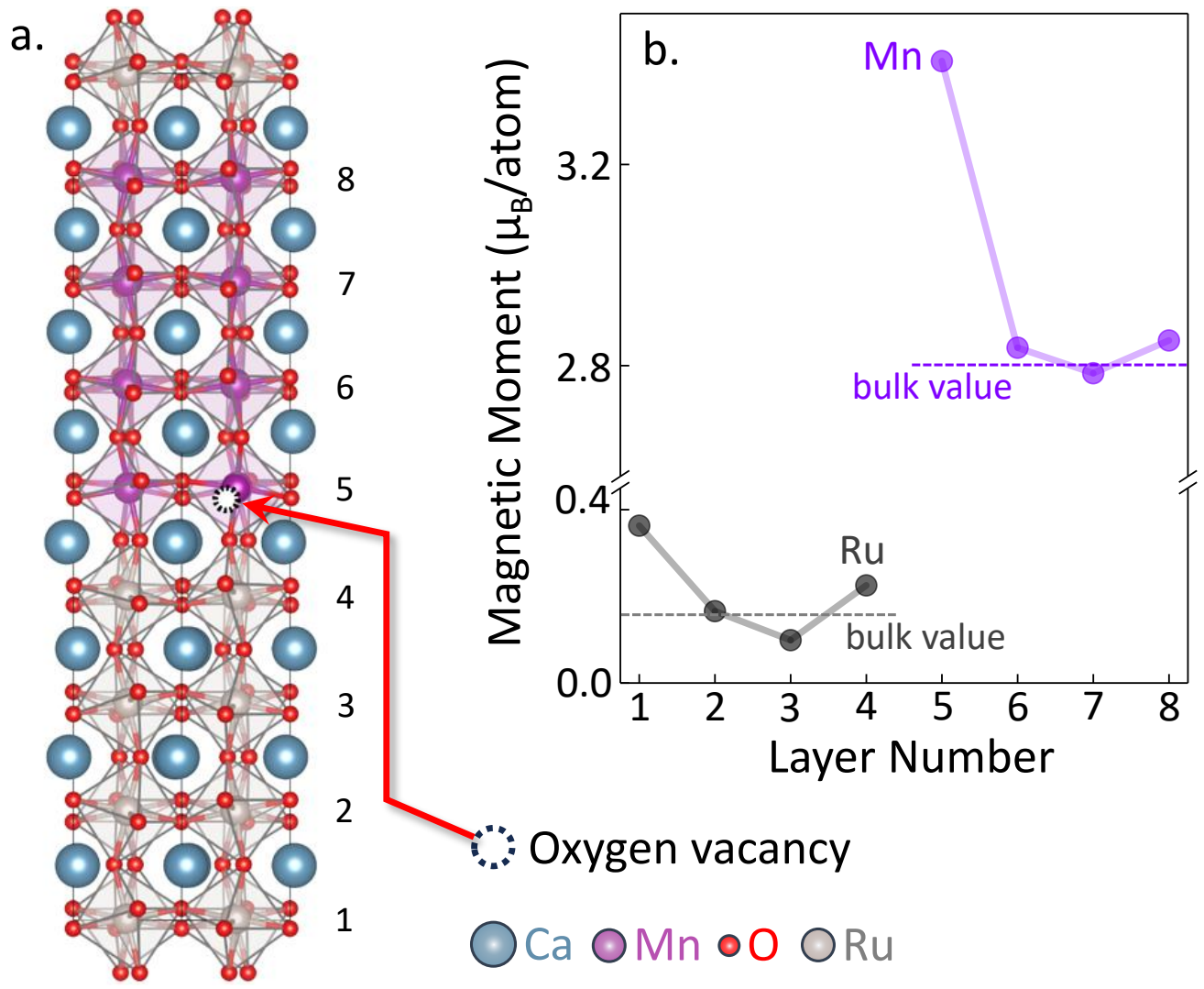

**Figure 4. a.** Crystal structure of the $CaRuO_3/CaMnO_3$ supercell with an oxygen vacancy introduced for one of the O atoms intermediate between the Mn and Ru atoms. **b.** Layer-resolved magnetic moments per atom exhibiting significant asymmetry between the top and bottom interfaces, with the increased magnetic moment in the interfacial $CaMnO_3$ layer that contains the vacancy. Dashed horizontal lines indicate the bulk-like values of the magnetic moments.

We note that the top and bottom interfaces in our supercells are identical by symmetry, and so our calculations using the nominal superlattice structure do not capture our measured asymmetry between the magnetism of the top and bottom $CaMnO_3$ interfacial layers. To explore the possible role of defect chemistry in causing this asymmetry, we next repeat our calculation procedure for a supercell containing oxygen vacancies at one interface. Specifically, we remove one of the four oxygen atoms intermediate between the Mn and Ru atoms at one of the interfaces, as shown in Figure 4a. Our resulting calculated magnetic moment per transition metal ion is substantially increased in the interfacial $CaMnO_3$ layer that contains the vacancy (see Figure 4b) pointing to a difference in the point defect chemistry, which could be introduced during the growth process, as the possible origin of the different sizes of the ferromagnetic moments at the two interfaces.

To rule out other possible origins of the observed magnetic asymmetry between the top and bottom $CaMnO_3$ interfacial layers, we repeated our calculation procedure for supercells with several plausible deviations from the structure shown previously in Fig. 3a. Specifically, we considered structures with mixed [001] and [110] oxygen octahedral tilt patterns leading to frustrated octahedral tilt connectivity at the interface, as well as superlattices with odd numbers of primitive cubic unit cell layers of $CaMnO_3$ and $CaRuO_3$. In each case, our calculations showed no significant magnetic asymmetry between the top and bottom interfacial $CaMnO_3$ layers (Figure S6 in the Supporting Information).

## CONCLUSION

In summary, we have discovered that the emergent ferromagnetism in $CaMnO_3$/$CaRuO_3$ oxide superlattices extends beyond the interfacial layer and presents an asymmetric distribution, challenging existing theoretical frameworks and suggesting a more complex interfacial behavior than previously recognized. Density functional calculations indicate that this ferromagnetism is driven by a double exchange mechanism, attributed to charge transfer from Ru to Mn ions, with defect chemistry – especially oxygen vacancies – likely playing a key role in creating the magnetic asymmetry observed at the interfaces. By pushing the boundaries of traditional magnetic interface studies and providing a deeper and more detailed insight into the atomic-level interactions at these interfaces, this work paves the way for future innovations in magnetic storage and spintronics.

## ACKNOWLEDGEMENTS

A.X.G. and J.R.P. acknowledge support from the US Air Force Office of Scientific Research (AFOSR) under award number FA9550-23-1-0476. A.M.T. and N.A.S. were funded by the European Research Council under the European Union's Horizon 2020 research and innovation program project HERO (Grant No. 810451) and by the ETH Zürich. Calculations were performed at the Swiss National Supercomputing Centre under Projects No. s889 and No. eth3 and on the EULER cluster of ETH Zürich. J.C., M.T. and M.K. acknowledge the support by the U.S. Department of Energy, Office of Science, Office of Basic Energy Sciences under Award No. DE-SC0022160. C.K. and P.S acknowledge support from the U.S. Department of Energy, Office of Science, Office of Basic Energy Sciences, the Microelectronics Co-Design Research Program, under Contract No. DE-AC02-05-CH11231 (Codesign of Ultra-Low-Voltage Beyond CMOS Microelectronics). This research used resources of the Advanced Light Source, which is a DOE Office of Science User Facility under Contract No. DE-AC02-05CH11231.

## *Supporting Information*

# Depth-resolved profile of the interfacial ferromagnetism in $CaMnO_3/CaRuO_3$ superlattices

J. R. Paudel[1], A. Mansouri Tehrani[2], M. Terilli[3], M. Kareev[3], J. Grassi[1], R. K. Sah[1], L. Wu[3], V. N. Strocov[4], C. Klewe[5], P. Shafer[5], J. Chakhalian[3], N. A. Spaldin[2], and A. X. Gray[1]

[1] *Physics Department, Temple University, Philadelphia, Pennsylvania 19122, USA*

[2] *Materials Theory, ETH Zurich, Wolfgang-Pauli-Strasse 27, CH-8093 Zürich, Switzerland*

[3] *Department of Physics and Astronomy, Rutgers University, Piscataway, New Jersey 08854, USA*

[4] *Swiss Light Source, Paul Scherrer Institute, 5232 Villigen, Switzerland*

[5] *Advanced Light Source, Lawrence Berkeley National Laboratory, Berkeley, California 94720, USA*

**email: axgray@temple.edu*

**Figure S1: Laboratory-based X-ray Diffraction (XRD) and Reflectivity (XRR)**

Figure S1a (next page) shows the laboratory-based X-ray diffraction θ-2θ spectrum measured on the [4 u.c. $CaMnO_3$ / 4 u.c. $CaRuO_3$] ×10 superlattice. The (002) $LaAlO_3$ substrate peak appears at $2\theta = 48^o$, consistent with prior studies [1,2]. The $0^{th}$-order superlattice peak ($SL_0$) is obscured by the substrate peak. The two $1^{st}$-order superlattice peaks ($SL_{-1}$) are observed at the symmetric angular positions of $44.65^o$ and $51.40^o$, approximately $\pm 3.4^o$ from the $0^{th}$-order peak. The superlattice peaks exhibit shapes characteristic of high-quality single-crystalline superlattices. The spectrum shows the expected number ($P - 2 = 8$) of SL thickness fringes for $P = 10$ superlattice periods. Observation of the pronounced SL thickness fringes indicates that the superlattice layers are reasonably smooth.

Figure S1b shows the complementary laboratory-based angle-resolved XRR spectrum measured on the same superlattice. The data, characterized by distinct thickness fringes, indicate a high-quality interface. The best fit of the experimental spectrum, obtained using X-ray optical code, yields a total thickness for the superlattice of approximately 28.05 nm, which is quantitatively consistent with the synchrotron-based XRR results presented in the main text.

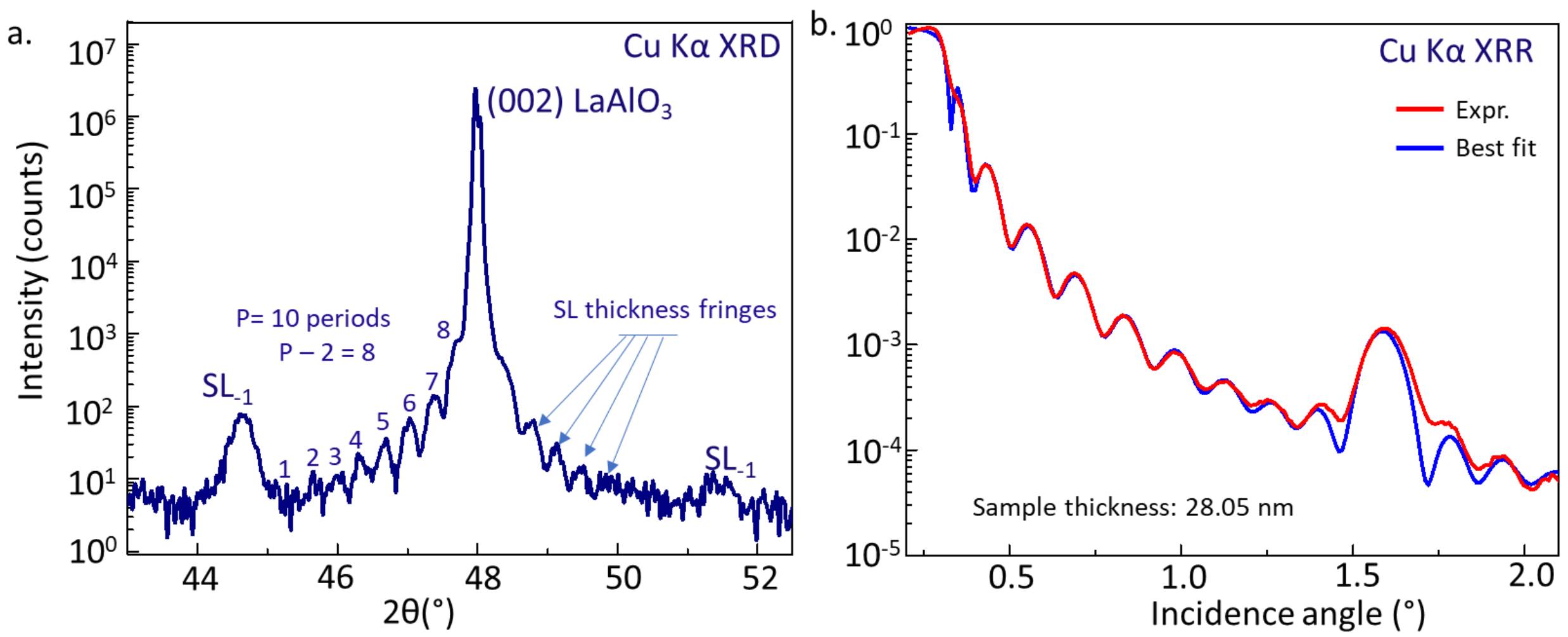

**Figure S1 a.** The X-ray diffraction (XRD) θ-2θ spectrum for the [4 u.c. $CaMnO_3$ / 4 u.c. $CaRuO_3$] ×10 superlattice. In this configuration, the $1^{st}$-order superlattice peaks ($SL_{-1}$) are prominently visible on either side of the substrate peak, appearing at 44.65° and 51.40°, respectively. The superlattice exhibits the anticipated number of SL thickness fringes, amounting to $P - 2 = 8$ for $P = 10$ superlattice periods. **b.** The X-ray reflectivity (XRR) measurement conducted on the same superlattice. The experimental XRR data, characterized by distinct thickness fringes, indicates a high-quality interface. The fitting of this data using the theoretical model yields a total thickness for the superlattice of approximately 28.05 nm.

**Figure S2: Synchrotron-based Standing-Wave X-ray Photoelectron Spectroscopy (SW-XPS)**

Figure S2 shows the results of the SW-XPS measurements carried out at the soft-X-ray ARPES endstation [3] of the high-resolution ADRESS beamline at the Swiss Light Source [4]. In the SW-XPS technique, shown schematically in Figure S2a, Ångstrom-level depth resolution is facilitated by generating an X-ray standing-wave (SW) interference field within a periodic superlattice sample [5,6]. Once the X-ray SW field is established within the sample, it can be translated vertically (perpendicular to the sample's surface) by approximately half of the superlattice period by scanning (rocking) the grazing X-ray incidence angle across the Bragg condition. Photoemission intensities of various core-level photoemission peaks are then recorded as a function of the X-ray incidence angle (rocking curves), thus facilitating element-specific depth profiling of the sample.

Figures S2b-d show experimental rocking-curve spectra for the integrated intensities of the Ca 2*p*, O 1*s*, and Ru 3*d* core-level peaks (red curves in the upper panels) and the best fits to the experimental data, calculated using an X-ray optical code [7,8] (blue curves in the lower panels). The extracted depth profile of the superlattice, yielding the values of the individual layer thicknesses and interface roughness (interdiffusion), is shown in panel e. The results are in quantitative agreement with both the lab-based XRR measurements shown in Supporting Figure S1 and the synchrotron-based resonant and non-resonant XRR measurements presented in Figure 1 of the main text.

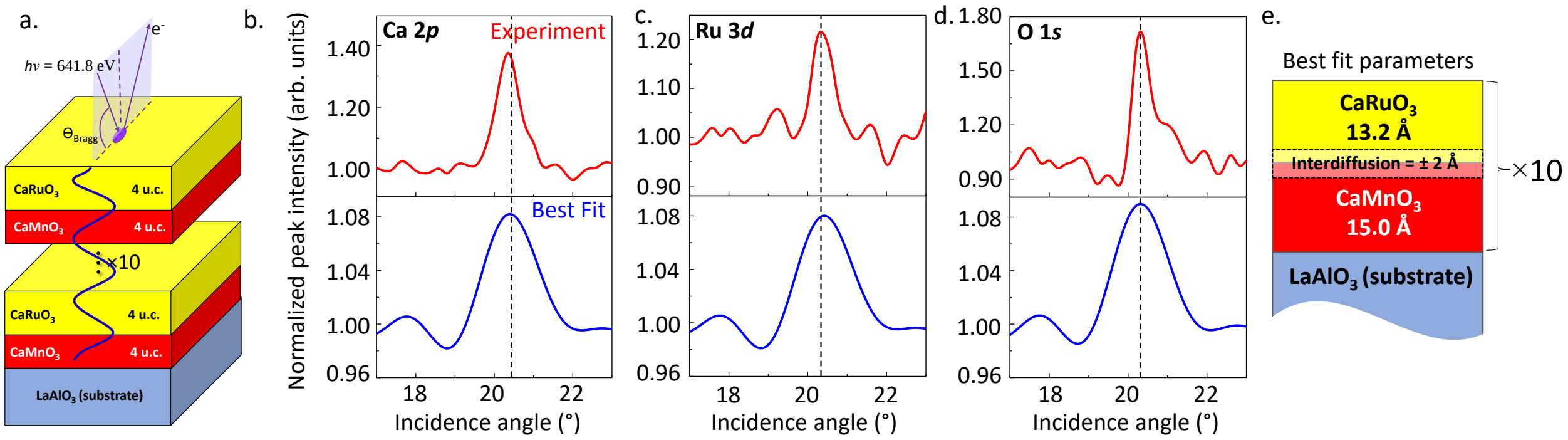

**Figure S2 a.** Schematic diagram of the standing-wave X-ray photoelectron spectroscopy (SW-XPS) experiment and the investigated superlattice structures consisting of 10 $CaMnO_3$/$CaRuO_3$ bilayers grown epitaxially on a $LaAlO_3$(001) substrate, with each bilayer consisting of 4 u.c. of $CaMnO_3$ and 4 u.c. of $CaRuO_3$. **b-d.** Best fits between the experimental (red curve, top panel) and calculated (blue curve, bottom panel) SW rocking curves for Ca 2*p*, O 1*s*, and Ru 3*d* core levels, respectively, for the superlattice. **e.** The resultant depth profile yielding the values of the individual layer thicknesses and interface roughness (interdiffusion).

**Figure S3: Hard X-ray Photoelectron Spectroscopy (HAXPES) Characterization**

The nominal chemical composition of the superlattices was confirmed using bulk-sensitive HAXPES measurements carried out using a laboratory-based spectrometer equipped with a 5.41 keV monochromated X-ray source and a Scienta Omicron EW4000 high-energy hemispherical analyzer. Figure S3 below shows wide-energy range HAXPES survey spectra for the $CaRuO_3$-terminated (red line) and $CaMnO_3$-terminated (blue line) superlattices. The presence of all expected elements (Ca, Mn, O, Ru, and C from the surface-adsorbed contaminant C/O layer) is confirmed by the presence of corresponding core-level peaks. Measurements were carried out at room temperature.

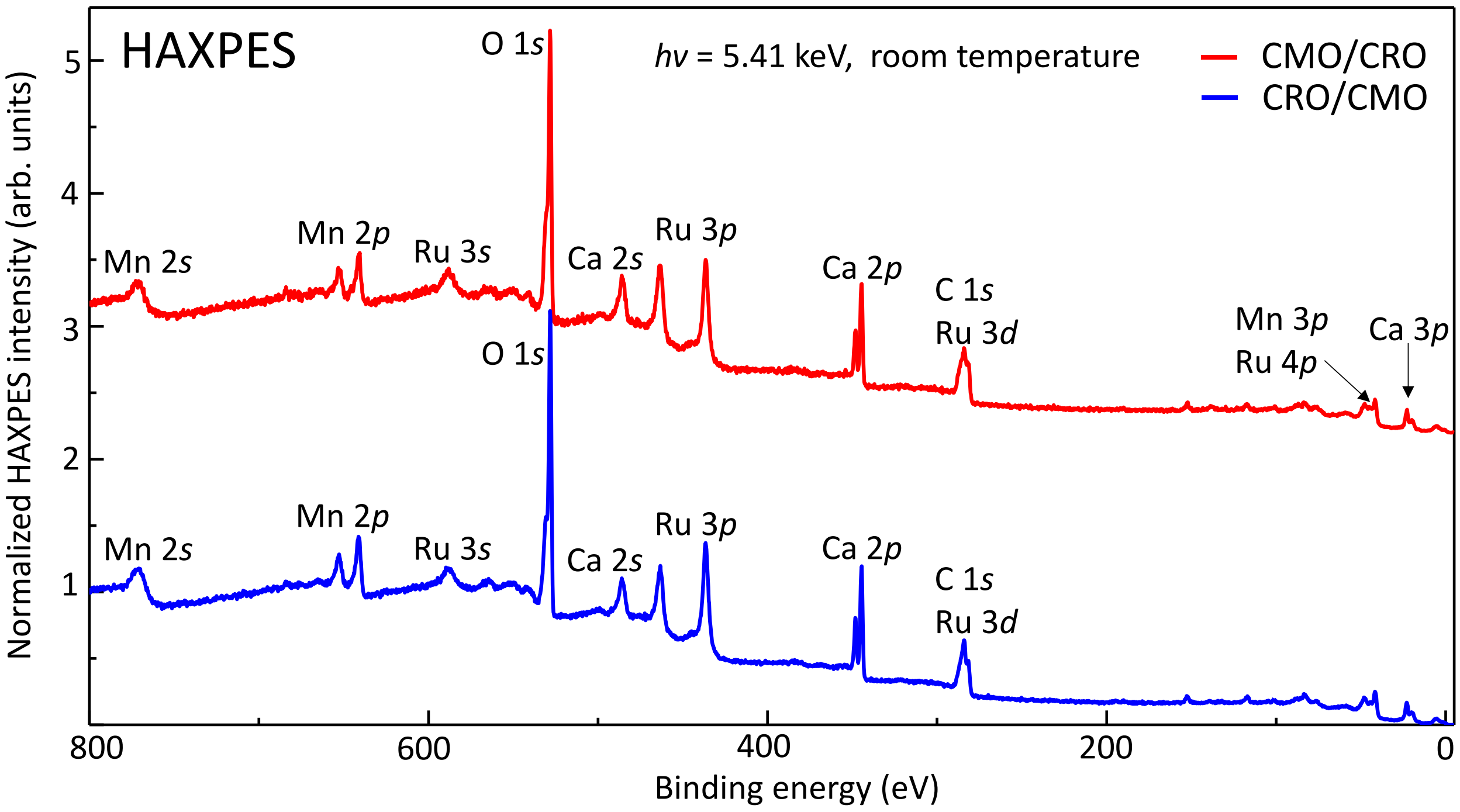

**Figure S3.** Wide-energy-range HAXPES survey spectra for the $CaRuO_3$ terminated (CMO/CRO) (red line) and $CaMnO_3$ terminated (CRO/CMO) (blue line) superlattices.

## Figure S4: X-ray Absorption Spectroscopy (XAS) in the Total Electron Yield (TEY) Mode

In order to directly probe the magnetic moment at the two types of interfaces ($CaMnO_3$/$CaRuO_3$ and $CaRuO_3$/$CaMnO_3$) in the superlattice, we conducted a comparative study using XAS/XMCD in the total electron yield (TEY) detection mode. TEY is a surface-sensitive XAS modality with an average probing depth of 2-5 nm, which decays exponentially from the surface into the bulk due to the short electron mean free path in solids [9,10].

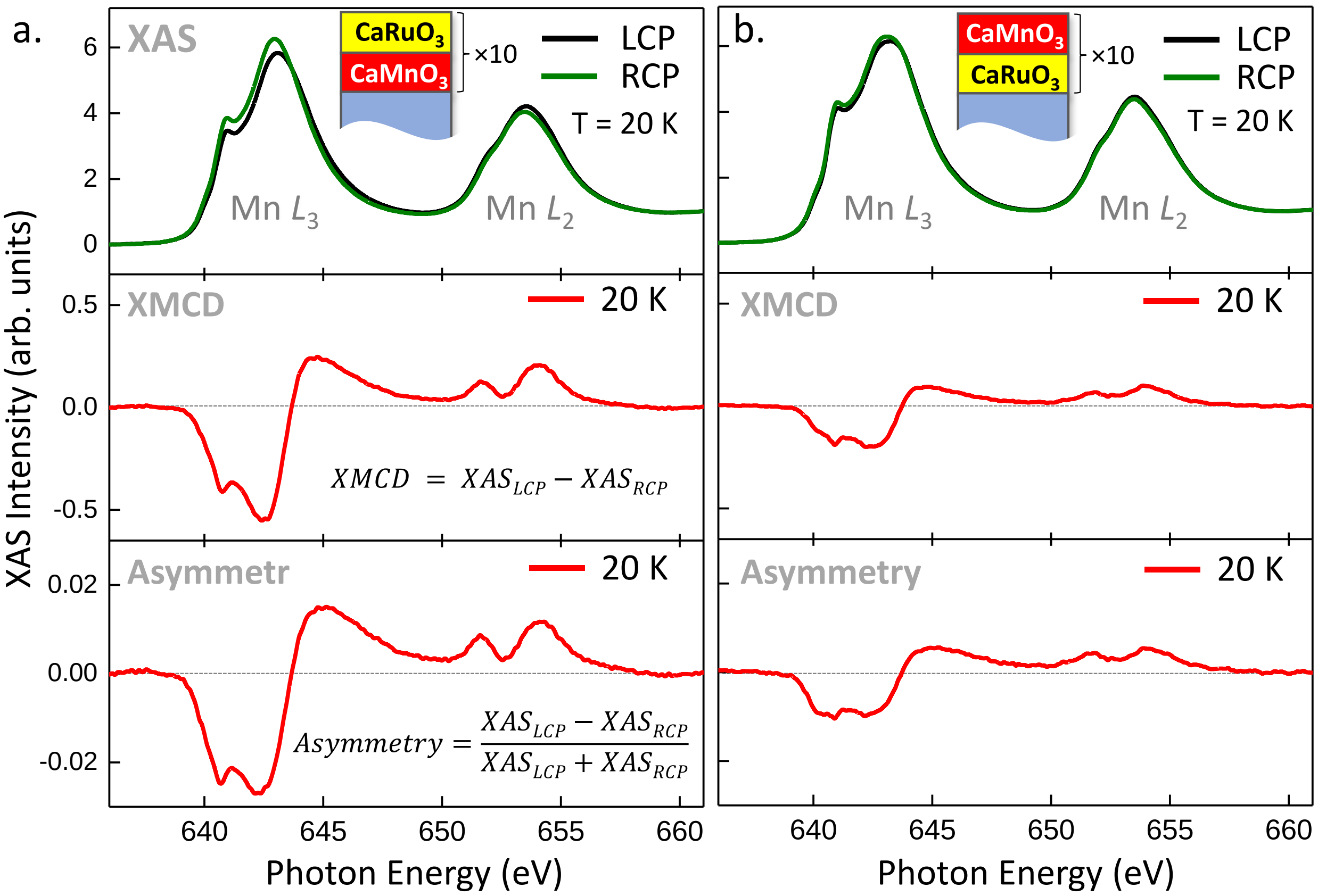

**Figure S4 a.** Circular polarization-dependent Mn $L_{2,3}$ X-ray absorption spectroscopy (XAS) spectra measured at T = 20 K (~80 K below $T_c$) on a [$CaMnO_3$/$CaRuO_3$]×10 superlattice. XMCD and asymmetry plots show strong magnetic signal. **b.** Similar measurements carried out on a [$CaRuO_3$/$CaMnO_3$] ×10 superlattice reveal significant suppression of the magnetic signal (by a factor of ~2.7).

An additional superlattice sample was grown in the same batch as the original superlattice, terminated with a $CaMnO_3$ layer instead of $CaRuO_3$. The XAS/XMCD measurements of the original $CaMnO_3/CaRuO_3$ sample (terminated with $CaRuO_3$) shown in panel **a.** are most sensitive to the $CaMnO_3/CaRuO_3$ ('top' type) interface. Conversely, a similar measurement of the $CaRuO_3/CaMnO_3$ sample (terminated with $CaMnO_3$) shown in panel **b.** is most sensitive to the $CaRuO_3/CaMnO_3$ ('bottom' type) interface. The measurements reveal a significantly weaker (×2.7) magnetic signal for the $CaRuO_3/CaMnO_3$ ('bottom' type) interface, qualitatively consistent with our $q_z$-dependent XRR measurements shown in Figure 2b of the main text.

**Figure S5: Symmetric magnetic interfaces – best fits for the constrained models**

To verify the observed difference (asymmetry) in the magnitudes of the Mn magnetic moments between the top and bottom $CaMnO_3$ interfaces, we repeated the fitting of the $q_z$-dependent XMCD asymmetry spectra at the Mn $L_3$ and $L_2$ edges with an additional constraint, requiring the magnitudes of $\Delta\beta_m$ to be the same for both interfaces while varying together in the model. The results of these calculations are shown in Figures S5a (Mn $L_3$ edge) and S5b (Mn $L_2$ edge). This minor modification of the model, affecting only one variable parameter, notably reduced the quality of the fit, resulting in major inconsistencies in the intensities of several XMCD peaks (labeled A, B, C, D, and E) as well as notable shifts in peaks C and F.

Most importantly, however, the best fits for the Mn $L_3$ and $L_2$ XRR XMCD data yielded different/inconsistent (and non-physical) values for the thickness of the magnetic layer. These additional calculations underscore the importance of a self-consistent analysis of the Mn $L_3$ and $L_2$ spectra, as shown in Figure 2 of the main text.

As an additional check, we repeated the calculations with the thickness of the two symmetric magnetic interfaces fixed in the model to the best-fit value obtained in the original calculation shown in Figure 2 of the main text. However, this approach produced an even worse fit for the Mn $L_3$ edge data and did not converge at all for the Mn $L_2$ edge fit, as depicted in Figures S5c and S5d below.

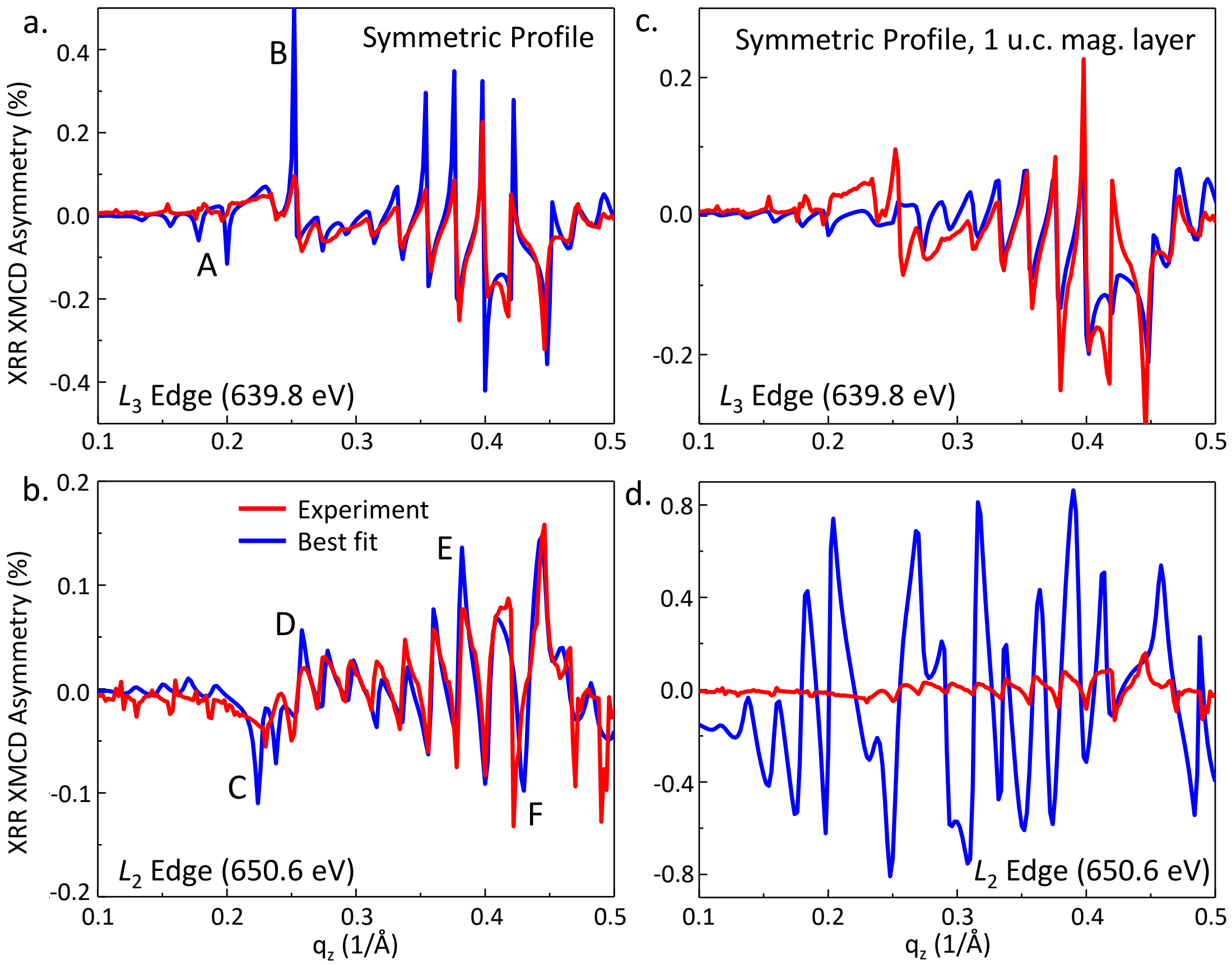

**Figure S5. a-b.** $q_z$-dependent XMCD asymmetry spectra (red curves) and the best-fit results to the experimental data (blue curves) for the Mn $L_3$ (panel a) and $L_2$ (panel b) absorption edges. The fitting was carried out with a constraint that required the magnitudes of $\Delta\beta_m$ to be the same for both interfaces, while being optimized together during the fitting process, along with variation in magnetic layer thickness. **c-d.** Additional fits, repeated with the thickness of the two symmetric magnetic interfaces constrained (fixed) in the model to the best-fit value obtained in the original calculation shown in Figure 2 of the main text.

**Figure S6: Possible deviations from the nominal interface structure and the resultant Mn magnetic moments**

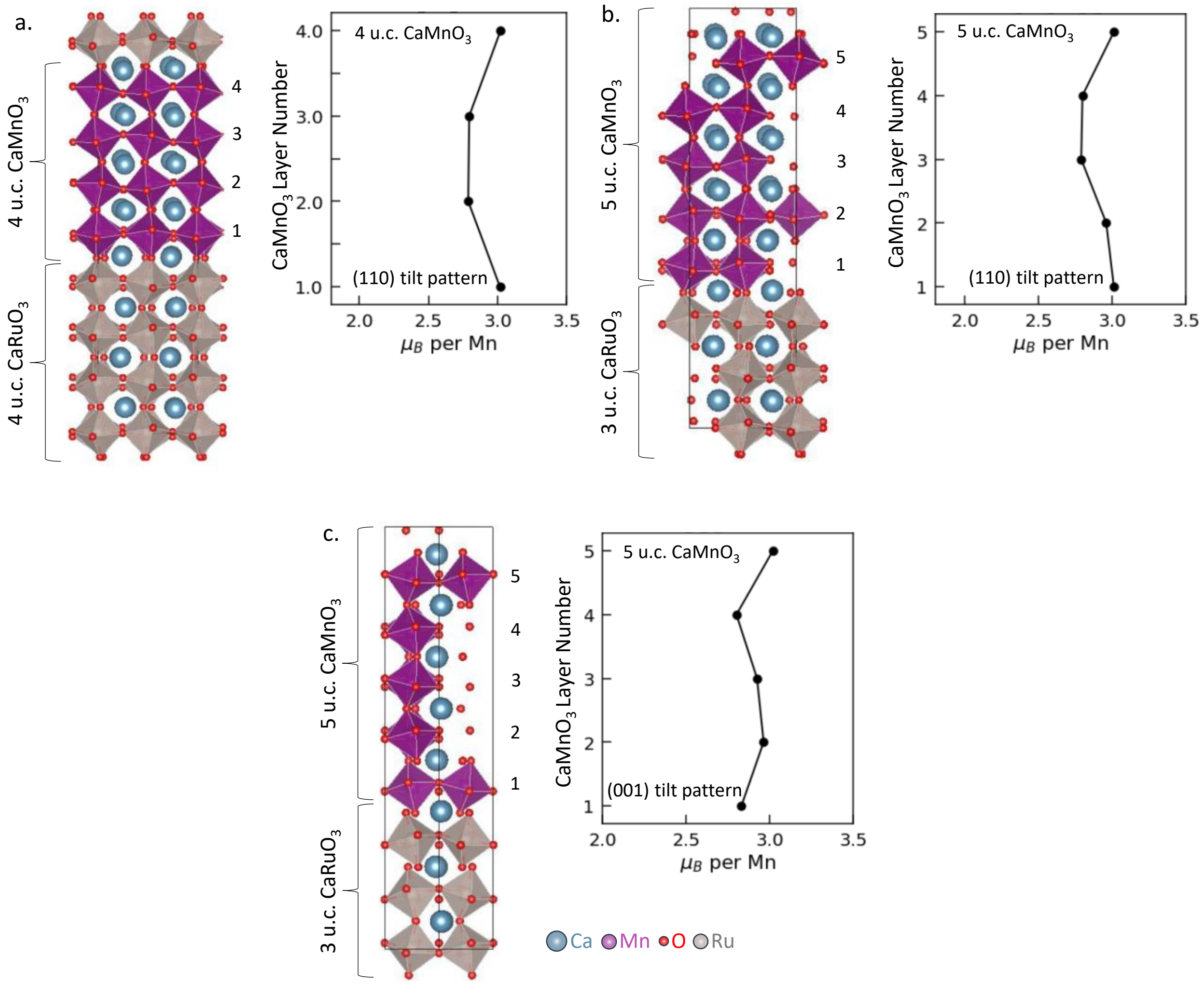

**Figure S6.** **a.** Crystal structure of the 4 u.c. / 4 u.c. superlattice with [110] oxygen octahedral tilt patterns in the interfacial $CaMnO_3$ layer and the resultant layer-resolved Mn magnetic moments. **b.** Similar calculations for the 3 u.c. / 5 u.c. superlattice. c. Similar calculations for the 3 u.c. / 5 u.c. superlattice with nominal [001] oxygen octahedral tilt patterns in the interfacial $CaMnO_3$ layer.

To investigate or rule-out other possible origins of the observed magnetic asymmetry between the top and bottom $CaMnO_3$ interfacial layers, we repeated our calculation procedure for supercells with several plausible deviations from the nominal structure shown previously in Figure 3a. Thus, in Figure S6a, we consider the nominal 4 u.c./4 u.c. superlattice structure, but with [110] oxygen octahedral tilt patterns in the interfacial $CaMnO_3$ layer. The right panel shows the resultant calculated layer-resolved Mn magnetic moments in the $CaMnO_3$ layer. Interestingly, we observed no significant magnetic asymmetry between the top and bottom interfacial $CaMnO_3$ layers.

We then repeated our calculation for a similar structure, but with an odd number of unit cells (5 u.c. and 3 u.c., respectively) of $CaMnO_3$ and $CaRuO_3$. The proposed structure and the corresponding Mn magnetic moment calculations are shown in Figure S6b. An additional calculation for this structure, but with [001] oxygen octahedral tilt patterns in the interfacial $CaMnO_3$ layer, are shown in Figure S6c. In each case, our calculations showed no significant magnetic asymmetry between the top and bottom interfacial $CaMnO_3$ layers.

**Details of the DFT Calculations**

We used the VASP DFT implementation [11] with the LDA exchange correlation functional, combined with a Hubbard U of 3 eV applied to the Mn 3*d* orbitals using the Dudarev approach [12], and the default VASP pseudopotentials with the following electrons included in the valence manifold: Ca $3s^2$, $3p^6$, $4s^2$; Mn $3p^6$, $4s^2$, $3d^5$; O $2s^2$, $2p^4$; and Ru $4p^6$, $4d^7$, $5s^2$. Our superlattices were constructed of four layers of $CaMnO_3$ alternating with four layers of $CaRuO_3$ (see Figure 3a in the main text), corresponding to both [001] and [110] growth directions, as well as mixed heterostructures with alternating [001]- and [110]-oriented slabs. We used an in-plane periodically repeated unit cell of size $\sqrt{2} \times \sqrt{2}$ times the simple cubic perovskite cell to accommodate rotations and tilting of the oxygen octahedra, with the in-plane lattice parameters set to the experimental lattice parameter of $LaAlO_3$. The *c* lattice parameter and the internal coordinates were adjusted to their lowest energy values using an energy cutoff of 800 eV, a *k*-point mesh of 12×12×1 for the smallest (80 atom) supercell adjusted for larger supercells and convergence criteria of $1\times10^{-7}$ eV and $1\times10^{-3}$ eV/A for the electronic and ionic relaxations, respectively. For the internal coordinate relaxations, numerous starting structures with different likely octahedral tilt patterns were used to ensure a thorough sampling of the structural landscape, and in all cases a range of likely magnetic orderings was explored.